\documentclass[reprint,superscriptaddress,amsmath,amssymb,aps,prL,twocolumn]{revtex4-1}

\pdfoutput=1

\usepackage{graphicx}

\usepackage{xcolor}

\usepackage{hyperref}
\hypersetup{
    colorlinks=true,
    linkcolor={blue!50!black},
    citecolor={blue!50!black},
    urlcolor={blue!80!black}
}

\makeatletter

\def\@fnsymbol#1{\ensuremath{\ifcase#1\or *\or *\or *\or
   *\or *\or * \or * \or *
   \or \ddagger\ddagger \else\@ctrerr\fi}}

\renewcommand{\fnum@figure}{\textsf{\textbf{Figure} \textbf{\thefigure}}}

\makeatother

\setcitestyle{super}

\begin{document}

\title{In situ Electric Field Skyrmion Creation in Magnetoelectric $\rm Cu_2OSeO_3$}

\noaffiliation

\author{Ping Huang}
\email{ping.huang@epfl.ch}
\affiliation{Laboratory for Quantum Magnetism (LQM), Institute of Physics, \'{E}cole Polytechnique F\'{e}d\'{e}rale de Lausanne (EPFL), CH-1015 Lausanne, Switzerland}
\affiliation{Laboratory for Ultrafast Microscopy and Electron Scattering (LUMES), Institute of Physics, \'{E}cole Polytechnique F\'{e}d\'{e}rale de Lausanne (EPFL), CH-1015 Lausanne, Switzerland}

\author{Marco Cantoni}
\affiliation{Centre Interdisciplinaire de Microscopie \'{E}lectronique (CIME), \'{E}cole Polytechnique F\'{e}d\'{e}rale de Lausanne (EPFL), CH-1015 Lausanne, Switzerland}

\author{Alex Kruchkov}
\affiliation{Laboratory for Quantum Magnetism (LQM), Institute of Physics, \'{E}cole Polytechnique F\'{e}d\'{e}rale de Lausanne (EPFL), CH-1015 Lausanne, Switzerland}

\author{Rajeswari Jayaraman}
\affiliation{Laboratory for Ultrafast Microscopy and Electron Scattering (LUMES), Institute of Physics, \'{E}cole Polytechnique F\'{e}d\'{e}rale de Lausanne (EPFL), CH-1015 Lausanne, Switzerland}

\author{Arnaud Magrez}
\affiliation{Crystal Growth Facility, Institute of Physics,  \'{E}cole Polytechnique F\'{e}d\'{e}rale de Lausanne (EPFL), CH-1015 Lausanne, Switzerland}

\author{Fabrizio Carbone}
\email{fabrizio.carbone@epfl.ch}
\affiliation{Laboratory for Ultrafast Microscopy and Electron Scattering (LUMES), Institute of Physics, \'{E}cole Polytechnique F\'{e}d\'{e}rale de Lausanne (EPFL), CH-1015 Lausanne, Switzerland}

\author{Henrik M. R{\o}nnow}
\email{henrik.ronnow@epfl.ch}
\affiliation{Laboratory for Quantum Magnetism (LQM), Institute of Physics, \'{E}cole Polytechnique F\'{e}d\'{e}rale de Lausanne (EPFL), CH-1015 Lausanne, Switzerland}

\date{\today}

\maketitle

\textbf{Magnetic skyrmions are localized nanometric spin textures with quantized winding numbers as the topological invariant\cite{braun_topological_2012}. Rapidly increasing attention has been paid to the investigations of skyrmions since their experimental discovery in 2009\cite{muhlbauer_skyrmion_2009}, due both to the fundamental properties and the promising potential in spintronics based applications\cite{fert_skyrmions_2013, nagaosa_topological_2013}. However, controlled creation of skyrmions remains a pivotal challenge towards technological applications. Here, we report that skyrmions can be created \textit{locally} by electric field in the magnetoelectric helimagnet $\mathbf{Cu_2OSeO_3}$. Using Lorentz transmission electron microscopy, we successfully write skyrmions \textit{in situ} from a helical spin background. Our discovery is highly coveted since it implies that skyrmionics can be integrated into contemporary field effect transistor based electronic technology, where very low energy dissipation can be achieved, and hence realizes a large step forward to its practical applications.}

To date, several categories of materials have been discovered to host skyrmions, including heavy metal/ferromagnet interfacial systems\cite{fert_skyrmions_2013}; metallic chiral alloys such as MnSi\setcitestyle{numbers,notesep={,},open={},close={}}(refs.~\cite{muhlbauer_skyrmion_2009, jonietz_spin_2010}), FeGe (ref.~\onlinecite{yu_near_2011}), $\rm Fe_xCo_{1-x}Si$ (ref.~\onlinecite{yu_real-space_2010}) and $\beta$-Mn-type Co-Zn-Mn (ref.~\onlinecite{tokunaga_new_2015}); and insulators $\rm GaV_4S_8$ (ref.~\onlinecite{kezsmarki_neel-type_2015}) and $\rm Cu_2OSeO_3$ \setcitestyle{numbers,notesep={,},open={},close={}}(refs.~\cite{seki_observation_2012, seki_magnetoelectric_2012}). As potential candidates of the building block for novel spintronic applications, efforts have been made to achieve creation and/or manipulation of skyrmions in these systems. Combination of electric currents and temperature gradients can rotate the skyrmion lattice (SkL) in MnSi (ref.~\onlinecite{jonietz_spin_2010}). Meta-stable skyrmions have been observed to exist in much larger regions in the phase diagram than the equilibrium ones via controlled temperature-magnetic field paths in MnSi (ref.~\onlinecite{oike_interplay_2016}) and $\rm Co_8Zn_8Mn_4$ (ref.~\onlinecite{karube_robust_2016}). Individual laser pulses have also been demonstrated recently to induce the formation of metastable skyrmions in FeGe (ref.~\onlinecite{berruto_speed_2017}). In the presence of a ferromagnetic background, creation of skyrmions at the heavy metal/ferromagnet interfaces have been realized by in-plane inhomogeneous electric currents\setcitestyle{super}\cite{jiang_blowing_2015}, and out-of-plane spin polarized currents\cite{romming_writing_2013} or electric fields\cite{hsu_electric-field-driven_2016} ($E$-fields) applied through a STM tip.

However, most of the aforementioned attempts of creating and manipulating skyrmions used electric current as the controlling parameter. This is not ideal because, firstly it is energy dissipating, and more importantly it is not straightforward to integrate into modern electronic technology based on field effect transistors. The insulating skyrmion host $\rm Cu_2OSeO_3$, on the other hand, provides an alternative due to magnetoelectric coupling arising from spin-orbital effects\cite{bos_magnetoelectric_2008, seki_observation_2012, seki_magnetoelectric_2012, omrani_exploration_2014}. As a result, emergent electric polarizations appear in the magnetically ordered phases of $\rm Cu_2OSeO_3$. The coupling between the external $E$-field and the emergent electric polarization (the $-\mathbf{E} \cdot \mathbf{P}$ coupling) in $\rm Cu_2OSeO_3$ provides a direct handle for tuning the relative energies of the competing magnetic textures, which enables the manipulation of skyrmions. For example, it has been demonstrated that the SkL in $\rm Cu_2OSeO_3$ can be rotated by an $E$-field\cite{white_electric-field-induced_2014}.

An extremely interesting consequence of the $-\mathbf{E} \cdot \mathbf{P}$ coupling is the possibility of creating skyrmions. Controlling of the size of the skyrmion phase pocket in $\rm Cu_2OSeO_3$ by $E$-field has recently been observed by magnetic susceptibility and microwave spectroscopy\cite{okamura_transition_2016} and confirmed by neutron scattering\cite{kruchkov_direct_2017}, yet local and real space/time demonstration has been lacking. Meanwhile, theoretical work of creating skyrmions in chiral crystal systems by $E$-field remains scarce and is limited to creating skyrmions from the ferromagnetic state\cite{mochizuki_creation_2016, mochizuki_writing_2015}, rather than from the low-energy spin helical phase.

\begin{figure*}[!htb]
 \centering
 \includegraphics{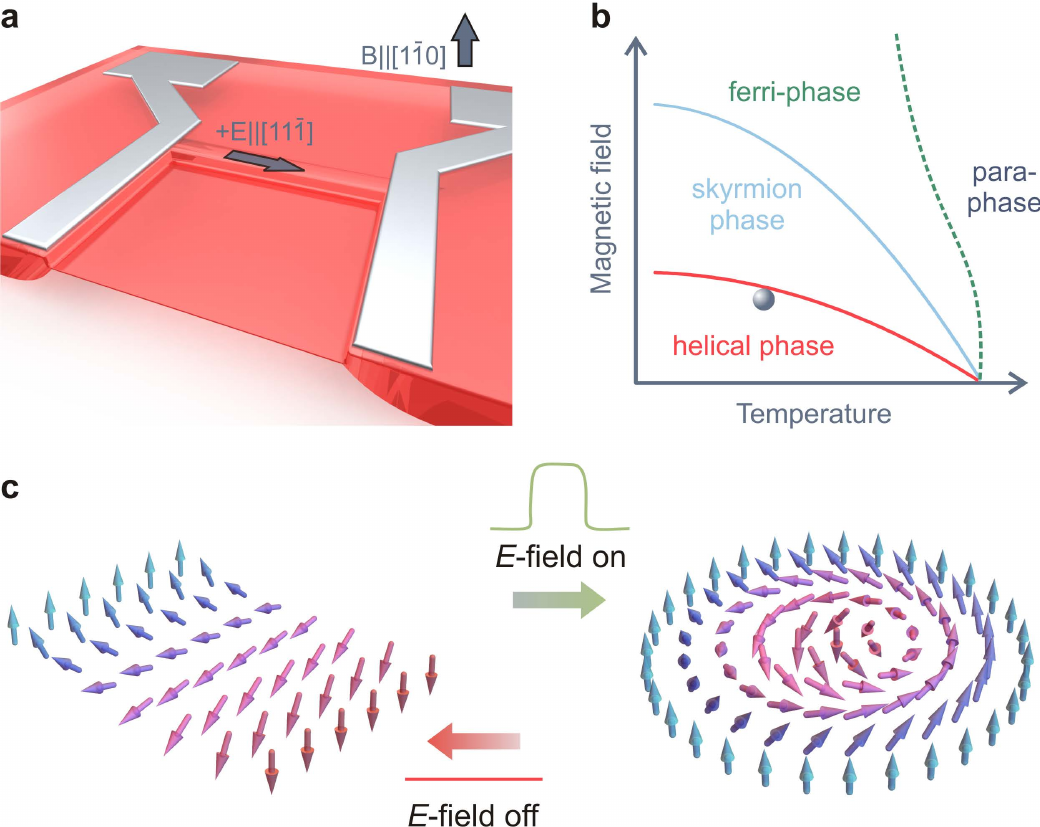}
 \caption{\textsf{\textbf{Illustrations of the sample configuration, the phase diagram and the transitions upon switching on and off the \emph{E}-field.} \textbf{a,} Schematic illustration of the H-bar LTEM sample configuration with Pt electrodes deposited by FIB directly on the surface of the sample. The 150 nm thick LTEM lamella is perpendicular to $\rm \mathsf{ [1\bar{1}0]}$ along which the magnetic field is applied. The electric field is along $\rm \mathsf{ [11\bar{1}]}$ on the sample plane. \textbf{b,} Schematic phase diagram for a thin slab $\rm \mathsf{Cu_2OSeO_3}$ sample. The initial state of the sample is in the helical phase and close to the helical-skyrmion phase boundary, as indicated by the solid dot. \textbf{c,} Schematic illustration of the transition of the spin textures from the helical phase to the skyrmion phase upon the application of the E-field, and vice versa.}}
 \label{fig:fig1}
\end{figure*}

Here, we report an experimental realization of \textit{locally} creating skyrmions by $E$-field from the helical phase observed \textit{in situ} by Lorentz transmission electron microscopy (LTEM). LTEM is a very suitable tool for studying spin textures with in-plane long wavelength modulations. What makes it particularly exciting is the capability of the \textit{in situ} tuning of conditions such as the $E$-field, the magnetic field and the temperature, while observing the system in real space and real time. Analysis of LTEM data can further give local and temporal information\cite{rajeswari_filming_2015}.

For the experiments reported here, a $\rm Cu_2OSeO_3$ single crystal was cut and mechanically polished to a slice 20 $\mu$m thick. Platinum electrodes for applying an in-plane $E$-field were deposited directly on the top surface of the sample by focused ion beam (FIB) with a spacing of 56 $\mu$m. A 20 $\mu$m wide section in between the electrodes was further milled down by FIB to a thickness of 150 nm to allow LTEM observation. Previous susceptibility and neutron studies were performed in the geometry $ \rm \mathbf{B} \parallel \mathbf{E} \parallel [111] $. However, that geometry is not suitable for our LTEM experiments since the electrodes will have to be deposited on the path of the incident electron beam on both the top and the bottom sides of the sample. Instead, we chose a geometry of $ \rm \mathbf{B} \parallel [1\bar{1}0] $ and $ \rm \mathbf{E} \parallel [11\bar{1}] $. In this geometry, the spontaneous electric polarization of the skyrmions is thus along $\rm [00\bar{1}] $\setcitestyle{numbers}(ref.~\onlinecite{seki_magnetoelectric_2012})\setcitestyle{super}, which is on the sample plane and couples to the applied $E$-field. The sample configuration is illustrated in Fig.~\ref{fig:fig1}a. More details about the sample preparation and the experiments can be found in the Method section and the Supplementary Information.

\begin{figure*}[!htb]
 \centering
 \includegraphics{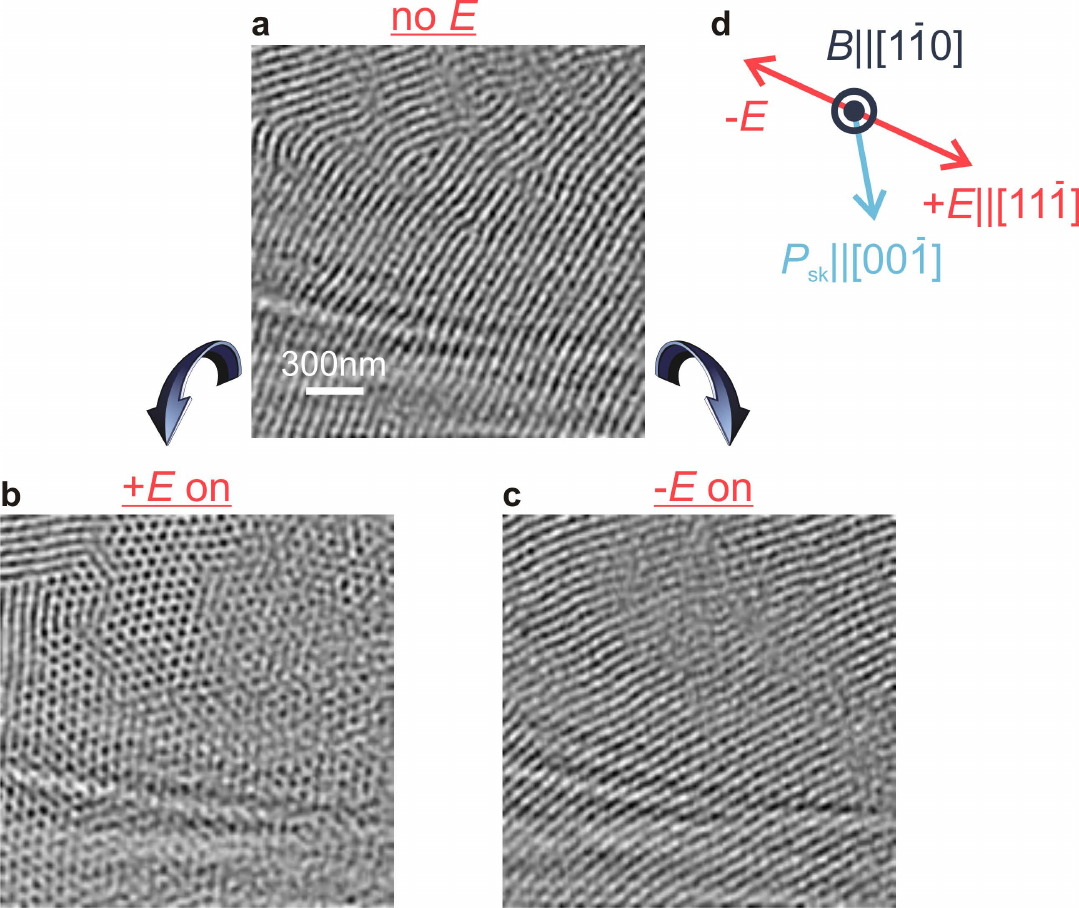}
 \caption{\textsf{\textbf{Real space LTEM images at the different \emph{E}-fields.} \textbf{a,} LTEM image with no applied \textit{E}-field. \textbf{b,} LTEM image at \textit{E}=+3.6 V/$\mu$m. \textbf{c,} LTEM image at \textit{E}=$-$3.6 V/$\mu$m. These real space observations clearly show that only the positive \textit{E}-field creates skyrmions. All figures are treated by a LoG filter. \textbf{d,} Directions of the \textit{E}-fields, the magnetic field and the spontaneous electric polarization of the skyrmion phase are shown with correspondence to the LTEM images.}}
 \label{fig:fig2}
\end{figure*}

In the phase diagram of nano-slab $\rm Cu_2OSeO_3$ samples, as sketched in Fig.~\ref{fig:fig1}b, the helical phase can be found at low magnetic fields, followed by the SkL phase until the ferrimagnetic phase is reached at high fields. Our measurements were performed at 24.7 K and 254 Oe, placing the sample in the helical phase but close to the SkL phase, as schematically indicated by the solid dot in the phase diagram in Fig.~\ref{fig:fig1}b. The application of an $E$-field along $[11\bar{1}]$ direction was observed to convert the helical magnetic texture into skyrmions, as schematically illustrated in Fig.~\ref{fig:fig1}c.

A typical LTEM image without the application of the $E$-field is shown in Fig.~\ref{fig:fig2}a. The system is in the helical phase with multiple helical domains having propagation vectors close to $ \rm [11\bar{1}] $ and $\rm [00\bar{1}]$ directions respectively. A few skyrmions can be found near the domain boundaries, demonstrating the proximity to the phase transition (More LTEM images with larger field of view can be found in the Supplementary Information). The existence of local skyrmions in the helical phase was recently addressed theoretically and experimentally\cite{muller_magnetic_2017}. After switching on a positive $E$-field, a large domain of skyrmions appears, as can be seen in Fig.~\ref{fig:fig2}b, demonstrating the creation of skyrmions by the $E$-field. The application of a negative $E$-field (Fig.~\ref{fig:fig2}c), on the other hand, does not show any sign of creating skyrmions, and only affected are some of the helical domains, for example, the bottom right part of Fig.~\ref{fig:fig2}c if compared with Fig.~\ref{fig:fig2}a at the same region. This asymmetric response, where only one $E$-field direction creates skyrmions, immediately rules out that the phase boundary is crossed by Joule heating due to the leakage current, which would have a symmetric $E$-field response\cite{oike_interplay_2016}. Therefore, it can be concluded that it is indeed the $E$-field effect which creates the skyrmions.

\begin{figure*}[!htb]
 \centering
 \includegraphics{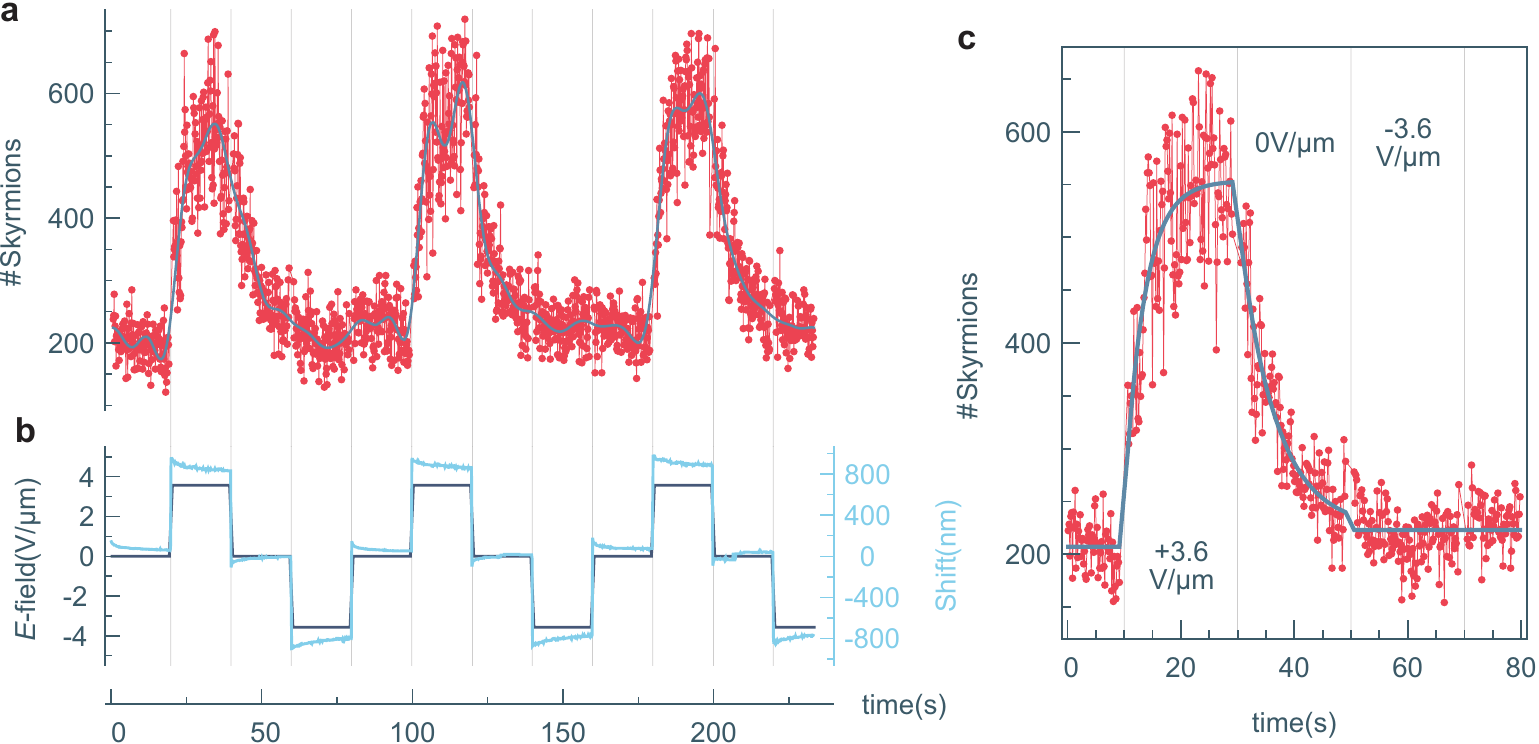}
 \caption{\textsf{\textbf{The statistics on the number of skyrmions.} \textbf{a,} The time dependence of the number of skyrmions in each image of the LTEM movie counted by the algorithm described in the main text. The solid line is a smoothing of the data for clarity. \textbf{b,} The applied bipolar \textit{E}-field sequence and the corresponding shifts of the LTEM images as a function of time, respectively. The step-like shape of the data in \textbf{a} together with its perfect correspondence to the \textit{E}-field profile in \textbf{b} fully demonstrate the mono-polar feature of creating skyrmions by \textit{E}-field. \textbf{c,} The average of the statistics into a single period of the applied \textit{E}-field. The solid lines are the fitting to the exponential growth/decay as described in the main text.}}
 \label{fig:fig3}
\end{figure*}

Furthermore, a sequence of $E$-fields, 0 V, $+$200 $V$ ($+3.6$ V/$\mu$m), 0 V and $-$200 V ($-$3.6 V/$\mu$m), were applied with each field lasting for 20 s, as shown in Fig.~\ref{fig:fig3}b, and during such $E$-field sequence, a LTEM movie was recorded with a frame interval of 150 ms. The applied in-plane $E$-fields caused horizontal shifts of the images, which allows for the direct determination of the directions, timings and relative strengthes of the $E$-fields. For $|E|=3.6$ V$/\mu$m, the shift is about 1,000 nm. The blue curve in Fig.~\ref{fig:fig3}b presents the shift as a function of time. It can be seen that the electric field reaches the image area immediately. The slight relaxation after each shift reflects weak screening due to the charging effect.

In order to quantify the number of skyrmions as a function of time throughout the acquired LTEM movies, we developed an algorithm capable of counting skyrmions coexisting with the helical phase. First, the LTEM images are filtered by a Laplacian of Gaussian (LoG) filter with the size corresponding to that of a skyrmion in the image. Then local minima are identified as candidate skyrmion positions. To distinguish actual skyrmions from local minima along a helical line, the orientational distribution maps of local intensity are calculated\cite{rao_computing_1989, rao_computerized_1992, hong_fingerprint_1998}. A local minima along a helical line will have the orientation parallel to its neighbours. Such candidates are therefore excluded. Manual inspection of selected frames confirms that the algorithm detects most of the skyrmions with a few false hits at helical domain walls and at bend contours in the images. Since imaging conditions stay constant during the movie, the false hits by the algorithm remain at a nearly constant background level of around 200 for all the images. More details on this algorithm, which was inspired by the algorithms used to identify fingerprints, can be found in the Supplementary Information.

Employing this algorithm makes it possible to extract the number of skyrmions for each of the 1,800 images in the LTEM movie. The number of skyrmions as a function of time is shown in Fig.~\ref{fig:fig3}a in the same time scale as in Fig.~\ref{fig:fig3}b where the applied $E$-field sequence is shown. The time evolution of the number of skyrmions shows repeated and reproducible writing of skyrmions by the positive $E$-fields. The process starts from $E=$0 V/$\mu$m, where the numbers of skyrmions are at the background level. An abrupt increase of the number of skyrmions is observed right at the time when switching on the $E$-field to $E=$+3.6 V/$\mu$m. Switching off the $E$-field results in a gradual monotonic decrease of the number of skyrmions and by the end of this $E$-field value, the number returns to the background level. No obvious change in the number of skyrmions can be observed after switching on an $E$-field in the opposite direction, $E=-3.6$ V/$\mu$m, consistent with the observation from the real space images that only the positive $E$-field creates skyrmions.

We averaged the counting results obtained from multiple $E$-field cycles into a single period, as shown in Fig.~\ref{fig:fig3}c, so that the dynamics of the $E$-field whole process can be further analyzed. The number of skyrmions after switching on and off the positive $E$-field can be described respectively by an exponential growth or decay: $N_i(t) = A_i(1-e^{-\frac{t-t_{i}}{\tau_i}}) + N_0$, where $t_i$ is the time points when the $E$-field is switched. The response time for creating skyrmions is shorter, $\tau_{write}=3.5$ s for switching on $E=$+3.6 V/$\mu$m. The decay of skyrmions after switching off the positive creating $E$-field is slower, $\tau_{persist}=6.4$ s (The full fitting results can be found in the Supplementary Information). This difference in time scales of the writing and relaxation processes implies different dynamics.

\begin{figure}[!htb]
 \centering
 \includegraphics{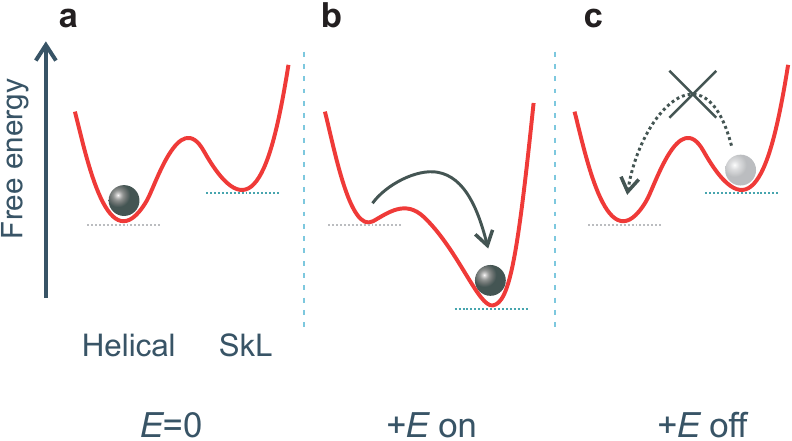}
 \caption{\textsf{\textbf{Illustration of the free energy profiles of the system near the helical-SkL phase boundary at different \textit{E}-fields.} The free energy profiles \textbf{a,} at no applied \textit{E}-field, \textbf{b,} when applying a creating \textit{E}-field, and \textbf{c,} after switching off the creating \textit{E}-field are illustrated respectively. The equilibrium states are indicated by the green dots while the meta-stable state by a faded one. The arrows indicate the change of the actual state of the system. The dotted lines in gray and green schematically indicate the energy levels of the helical phase and the SkL phase respectively in each situation.}}
 \label{fig:fig4}
\end{figure}

Qualitative insight to this phenomena can be achieved by considering the free energy profile as sketched in Fig.~\ref{fig:fig4}. Tuning the magnetic field inside the helical phase close to the skyrmion phase, the two states are almost degenerate in energy, with a barrier between them, as illustrated in Fig.~\ref{fig:fig4}a. The sample configuration we chose ensures that both the positive and the negative $E$-fields couple with the spontaneous polarizations of the skyrmions. Positive $E$-fields lower the energy of the skyrmion phase, which leads to the creation of skyrmions.

Furthermore, the separation of the time scales in the writing and relaxation processes suggests the possibility of potential barrier engineering so as to increase $\tau_{persist}$, \emph{e.g.} through pinning, while reducing $\tau_{write}$ through larger fields or thinner devices. This is because the effective barrier in the presence of a large positive electric field can be different and smaller (Fig.~\ref{fig:fig4}b) than the effective barrier in the absence of $E$-field (Fig.~\ref{fig:fig4}a and c).

In summary, we report the experimental realization of the writing of skyrmions in the magnetoelectric compound $\rm Cu_2OSeO_3$ conducted and observed \textit{in situ} by LTEM. Rather than starting from the spin polarized state mostly considered in theoretical proposals, we start from the helical phase. In our $ \rm \mathbf{B} \parallel [1\bar{1}0] $ and $ \rm \mathbf{E} \parallel [11\bar{1}] $ sample geometry, the transverse $E$-field is found to write skyrmions in only one polarity and to induce no obvious effects in the other one. Using an algorithm to count the number of skyrmions in each of the 1,800 frames of a LTEM movie, we extract respectively the reaction and the holding times of the $E$-field created skyrmions, which implies the possibility of engineering the effective energy barriers to achieve efficient writing of skyrmions by $E$-field while keeping information lossless even when powered off.

While in this demonstration experiment we prioritized a transverse $E$-field configuration for the LTEM investigation, the concept can readily be generalized to a grid pattern where arrays of electrodes arranged orthogonally on either side of a 50-100 nm thick slab would allow to apply longitudinal $E$-fields locally, hence forming an addressable array of skyrmion bits. We therefore believe that our observation opens the possibility for an alternative route towards skyrmionic memory than the so-called skyrmion race-track concept.

\section*{Methods}

High-quality $\rm Cu_2OSeO_3$ single crystals were synthesized by the method of chemical vapor transport redox reactions. A single crystal was aligned and cut so that the main plane was perpendicular to $[1 \bar{1} 0]$ direction. Then the rectangularly shaped sample was mechanically polished down to 20 $\mu$m. FIB was adopted to further mill a TEM lamella of 150 nm thick and 20 $\mu$m wide at the edge, forming a conventional H-bar configuration. Platinum electrodes were deposited by FIB directly on the upper surface of the sample 56 $\mu$m apart. A Gatan liquid helium sample holder with 4 feed-throughs was used to control the temperature of the sample as well as to connect to external instruments. All experiments were carried out on an FEI Titan Themis TEM in Lorentz mode. A Keithley 2400 source-meter controlled by Labview was used to apply dc voltages. Image data were treated and analyzed in Matlab.

\section*{Acknowledgments}
The authors would like to thank Dani\`{e}le Laub and Barbora B\'{a}rtov\'{a} for their help in sample fabrication. The work was supported by the Swiss National Science Foundation (SNSF) through project 166298, Ambizione Fellowship 168035, the Sinergia network 171003 for Nanoskyrmionics, and the National Center for Competence in Research 157956 on Molecular Ultrafast Science and Technology (NCCR MUST), as well as the ERC project CONQUEST.

\section*{Author Contributions}

H.M.R. organized the project, P.H., M.C. and H.M.R designed the experiment, A.M. synthesized the crystalline samples, P.H. and M.C performed the LTEM experiments, P.H. and H.M.R analyzed the data, all authors contributed to the interpretation of the data. P.H., H.M.R and F.C wrote the paper.

\section*{Competing financial interests}

The authors declare no competing financial interests.

\end{document}